\documentclass[format=acmsmall, review=false, screen=true]{acmart}

\usepackage{booktabs} 

\usepackage{multirow}
\usepackage{xcolor}
\definecolor{Gray}{gray}{0.80}
\usepackage{colortbl}
\usepackage{tabularx}
\usepackage{array} 
\newcolumntype{?}{!{\vrule width 1pt}}

\usepackage{hhline}

\usepackage[ruled]{algorithm2e} 

\SetAlFnt{\small}
\SetAlCapFnt{\small}
\SetAlCapNameFnt{\small}
\SetAlCapHSkip{0pt}
\IncMargin{-\parindent}

\acmJournal{PACMHCI}
\acmYear{2018} \acmVolume{2} \acmNumber{CSCW} \acmArticle{67} \acmMonth{11} \acmPrice{15.00}

\setcopyright{acmcopyright}

\acmDOI{10.1145/3274336}

\received{April 2018} 
\received[revised]{July 2018}
\received[accepted]{September 2018}

\begin{document}
\title[Who is Addressed in this Comment?]{Who is Addressed in this Comment? Automatically Classifying Meta-Comments in News Comments}

\author{Marlo H\"aring}
\orcid{}
\affiliation{%
  \institution{University of Hamburg, Germany}
  \streetaddress{Vogt-K\"olln-Stra{\ss}e 30}
  \city{Hamburg} 
  \state{Germany} 
  \postcode{22527}
}
\email{haering@informatik.uni-hamburg.de}

\author{Wiebke Loosen}
\affiliation{%
  \institution{Hans-Bredow-Institute, Germany}
  \streetaddress{Rothenbaumchaussee 36}
  \city{Hamburg} 
  \state{Germany} 
  \postcode{20148}
}
\email{w.loosen@hans-bredow-institut.de}

\author{Walid Maalej}
\affiliation{%
  \institution{University of Hamburg, Germany}
  \streetaddress{Vogt-K\"olln-Stra{\ss}e 30}
  \city{Hamburg} 
  \state{Germany} 
  \postcode{22527}
}
\email{maalej@informatik.uni-hamburg.de}

\renewcommand{\shortauthors}{M. H\"aring et al.}

\begin{abstract}
User comments have become an essential part of online journalism. However, newsrooms are often overwhelmed by the vast number of diverse comments, for which a manual analysis is barely feasible. Identifying meta-comments that address or mention newsrooms, individual journalists, or moderators and that may call for reactions is particularly critical. In this paper, we present an automated approach to identify and classify meta-comments. We compare comment classification based on manually extracted features with an end-to-end learning approach. We develop, optimize, and evaluate multiple classifiers on a comment dataset of the large German online newsroom SPIEGEL Online and the ``One Million Posts'' corpus of DER STANDARD, an Austrian newspaper. Both optimized classification approaches achieved encouraging $F_{0.5}$ values between 76\% and 91\%. We report on the most significant classification features with the results of a qualitative analysis and discuss how our work contributes to making participation in online journalism more constructive.
\end{abstract}

%
%



\maketitle

\section{Introduction}

It is becoming increasingly difficult for online newsrooms to handle the vast amount of user comments, which are heterogeneous in content and quality \cite{sood2012automatic}. For example, one of the most popular German online news sites, SPIEGEL Online, publishes $\sim$1.2 million user comments per year, which amounts to more than 3,000 comments per day and that is disregarding blocked comments and comments on social media. For community moderators, a manual selection of meaningful and highly qualified comments is neither easy nor scalable.
Journalists and journalism researchers repeatedly mention this problem: finding particularly useful or high-quality comments is like finding a needle in a haystack  \cite[p.387]{braun2011hosting} \cite{heise2014publikumsinklusion, reimer2015publikumsinklusion, park2016supporting}. Developing tools to assist moderators, journalists, and newsrooms to analyze, filter, and summarize user comments has been identified as a primary challenge for news organizations \cite{diplaris2012making, diakopoulos2015picking, diakopoulos2015editor}.

Research has shown that most journalists have a clear sense of what they deem useful user contributions \cite{loosen2017exploratory}. For instance, journalists particularly appreciate user feedback that reports errors in articles, include additional information on a topic, or contain critique addressed to the quality of an article. Media companies can use this information to improve journalistic work, correct articles, answer frequent questions, or gather feedback on the quality of their news coverage.

A previous study by Loosen et al. \cite{loosen2017exploratory} demonstrated, through group discussions with journalists and community-moderators, that the prospect of a software system for analyzing user comments was highly welcomed. One feature journalists considered particularly useful is the ability to identify the addressee in comments, for example, the newsroom or media organization, the author of the article being commented on, actors mentioned in the article, or other actors and users. This would help to direct comments to the newsroom or to single journalists that may call for reactions as correcting facts, answering questions, or providing additional information. This is all the more the case as it is also likely that user comments that address the author or the newsroom directly contain elements of media critique or praise \cite{craft2016reader}. 

Our work aims to develop and evaluate an approach to automatically identify and classify user comments based on whom they address. We focus on comments that are not (only) related to the article but address, for instance, the media company, a journalist, or a community-moderator. We call these comments ``meta-comments''.
The contribution of this paper is threefold. \textbf{First}, we empirically explore and evaluate the solution space for this classification task based on supervised and end-to-end machine learning approaches with respective hyperparameter optimization.
\textbf{Second}, we propose a neural network model for the end-to-end learning which outperforms state-of-the-art comment classification reported by Schabus et al.~\cite{Schabus2017}.
\textbf{Third}, we give insights into designing comment analytics tools and use cases for the information extracted from meta-comments.

The remainder of the paper is structured as follows. Section \ref{sec:study-design} introduces the research questions, method, and data. Section \ref{sec:dat-ana} describes the data analysis process and the training of different word and comment embedding models. Section \ref{sec:freatures-eng} outlines the analysis and deduction of machine learning features for a supervised machine learning approach. In Section \ref{sec:cla-exp-opt} we experiment with and compare the accuracy of an end-to-end learning approach with traditional machine learning based on manually extracted features. In Section \ref{sec:qualitative}, we use the classifier to classify unseen user comments and qualitatively analyze the results. We then discuss the threats to validity (Section \ref{sec:threats}), related work (Section \ref{sec:rel-work}), the implications of our findings (Section \ref{sec:discuss}), and conclude the paper in Section \ref{sec:conclusion}.

\section{Research Design}
\label{sec:study-design}

\subsection{Research Questions}

Following Neuberger \cite{neuberger2009internet}, we can differentiate between user comments related to the ``object level'' and those related to the ``meta level''. Comments at the object level refer to \textit{what} is covered, those at the meta level refer to \textit{how} something is covered by the newsroom or individual journalists. Actors mentioned or addressed within the object level are often prescribed through the topic of the respective article, for instance, politicians, companies, or celebrities. Comments addressing the writing performance or giving general feedback to the author of the article belong to the meta level. In this paper, we focus on the \textbf{meta-addressees}  and use a hierarchy inspired by Loosen et al.~\cite{loosen2013publikumsinklusion}:
\begin{itemize}
    \item[--] \textbf{Media}: covers the media companies, their editing, and news coverage, for instance, SPIEGEL Online (de), DER STANDARD (at), New York Times (us), or The Guardian (uk).
    \item[--] \textbf{Journalist}: refers to the article's author or other persons involved as editors or reporters.
    \item[--] \textbf{Community-Moderator}: refers to those who manage comment sections, read comments, actively participate in discussions, release, or block comments from the comment section.
\end{itemize}

Our goal is to identify whether a user comments is a \textbf{meta-comment} or not and then to classify meta-comments regarding their meta-addressees. A user comment is a meta-comment if it addresses at least one meta-addressee. We focus on three research questions:

\begin{itemize}
    \item [\textbf{RQ1}] Which classification approach/configuration is the most accurate for classifying meta-comments? 
    \item [\textbf{RQ2}] What are informative machine learning features among text features, semantic features, and comments' metadata to identify and classify meta-comments?
    \item [\textbf{RQ3}] Which information do classified meta-comments contain and how would it be useful?
\end{itemize}

\subsection{Research Method}
\begin{figure}[tbh]
  \includegraphics[width=\textwidth]{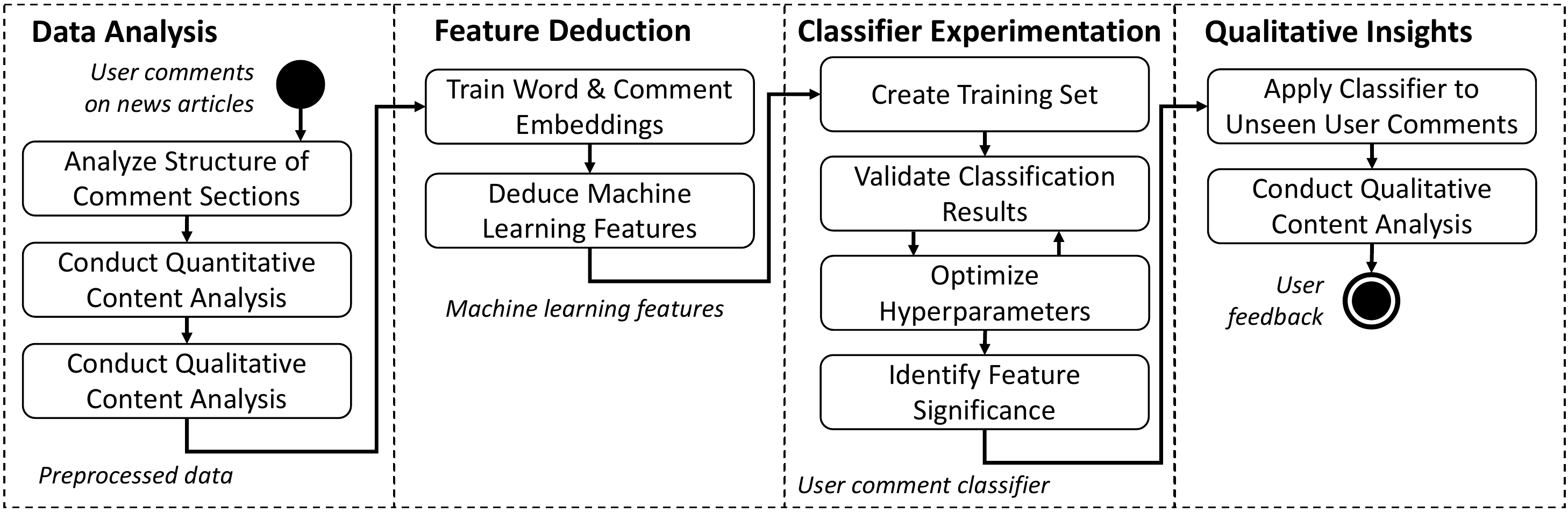}
  \caption{Overview of our research methodology with four main  consecutive steps.}
  \label{fig:research_steps}
\end{figure}

Figure \ref{fig:research_steps} shows an overview of our methodological framework, which comprises four consecutive phases.
To answer RQ1, we first deduced machine learning features for a supervised learning approach from a qualitative content analysis and related work. We trained the word and comment embeddings \cite{mikolov2013efficient, le2014distributed} for text features, semantic features \cite{rumelhart1988learning}, and for applying transfer learning \cite{michalski1983theory} on an end-to-end learning approach \cite{lecun2015deep}. We manually labeled a training set of user comments posted on SPIEGEL Online and combined it with the ``One Million Posts'' corpus to optimize the hyperparameter configuration for different classifiers and classification approaches. 
For RQ2, we calculated the most significant features for each meta-addressee class. 
For RQ3, we applied the trained classifier on a random subset of unlabeled user comments, read the classified comments, and qualitatively analyzed their content. 
The details of each step are discussed in the corresponding result section below.  

\subsection{Research Data}
To answer our research questions, we used two datasets: (1) user comments posted on SPIEGEL Online\footnote{http://www.spiegel.de/} (SPON) and (2) the ``One Million Posts'' (OMP) corpus  \cite{Schabus2017}. 
We selected the SPON news page for two reasons. First, SPON is the most-read online German newspaper according to Alexa.com \cite{alexa2017}. Second, the topics covered are diverse and structured in articles, forums, and comments. We collected a comprehensive sample of published user comments from 01-01-2000 to 28-02-2017 with their respective metadata and all archived articles and forums. The data collection took one week and we did not notice any changes of forum features between old and new forums. Our sample comprises 11,276,843 comments (with title, text, timestamp, username, department, and quoted comments if available), 515,522 articles (with title, introduction, text, date, and partly author names), and 181,399 forums (with title and department). Most SPON articles are signed by an acronym to state the author, while the acronyms are  assigned to full names in the imprint. However, we could only identify the full author names for 16\% of the news articles as many assignments were missing.

Additionally, we used the partly annotated comments of OMP, a dataset that consists of 11,773 labeled and one million unlabeled German online user comments posted on DER STANDARD, an Austrian newspaper website. The authors define the annotation category ``feedback'' as: ``Sometimes users ask questions or give feedback to the author of the article or the newspaper in general, which may require a reply/reaction'' \cite{Schabus2017}. This description is equivalent to our meta-comment definition.

\section{Data Analysis}
\label{sec:dat-ana}
We describe the structure of the comment sections, the quantitative, and qualitative content analysis.

\subsection{Structure of the Comment Sections}
\begin{figure}[hbt]
  \includegraphics[scale=.8]{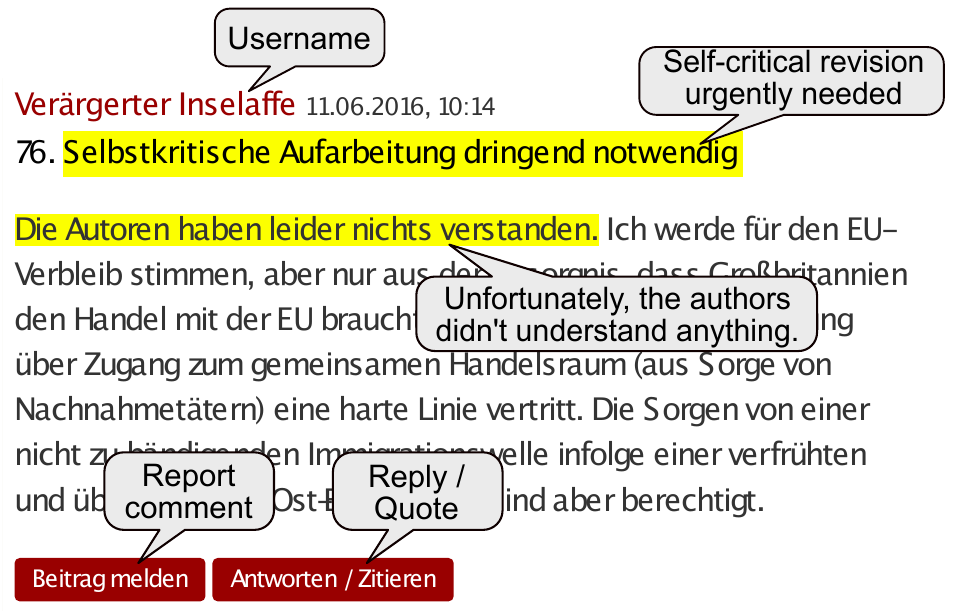}
  \caption{Example of a meta-comment in the SPON comment section.}
  \label{fig:spon_comment}
\end{figure}

SPON's comment section sorts user comments by time. It does not structure the comments in threads. Figure \ref{fig:spon_comment} shows an example of a SPON meta-comment. To post a comment on a news article, (1) users have to log in with either a SPON or Facebook account, (2) browse to the article's forum, and (3) compose a comment with a text and an optional title. Alternatively, users can ``Reply / Quote'' an existing user comment, which adds its text as a linked quote to the user comment. SPON forum moderators review each comment to check if it complies with the terms of use before it is publicly released on the SPON website. In our dataset, SPON forum moderators also contributed infrequently (1,216 comments) with the username ``sysop'' to the discussion \cite{spon2018}.

DER STANDARD's comment section structures comments into threads and users can rate existing user comments as ``worth reading'' or ``not worth reading''.
There are different filter and sort options. Users can filter the comments to see all postings, top postings, or postings by moderators and sort the comment list by date or rating. Forum moderators use their own name to write comments. They consider themselves as participants as opposed to rigid comment administrators and supplement the discussions through active participation, if they consider it beneficial to a discussion \cite{derstandard2018}.

\subsection{Quantitative Content Analysis}
We describe only the SPON dataset as Schabus et al.~\cite{Schabus2017} report on the OMP dataset in-depth. The number of SPON user comments per year has steadily increased from 2005 to 2011 from 0.1 million to 1.6 million. From 2011 to 2015, users posted between 1.2 and 1.6 million user comments per year. Users posted the majority of comments in the politics (4.5 million, 39.7\%) and economy sections (2.5 million, 21.9\%). The other leading sections are sport, panorama, culture, science, technology, life \& learning, car, health, career, and traveling. Each of them covers less than one million user comments in total (8.9\%). The average length of a comment's title is two words and 69 words for the text. 61\% of the comments contain a quote. The average number of words for the title of a SPON article is seven words, while the average length of an article text is 457 words. Users were able to comment on 32.8\% of all articles. On average, one forum (article) contains 66 user comments.

\subsection{Qualitative Content Analysis}
\label{sec:qual_cont_anal}
We conducted a qualitative content analysis of 1,000 randomly selected SPON user comments to better understand and quantify meta-comments and to identify potential useful machine learning features for our classification task. Each of the 1,000 comments was independently labeled by two human coders. We developed a coding guide for the labeling process in collaboration with communication researchers. It describes the labeling task with examples and defines each meta-addressee class to increase the quality of the manual labeling. Provided with a coding guide, student assistants labeled the comments. The coding guide and further resources are available on our project website\footnote{https://scan.informatik.uni-hamburg.de/user-comment-analysis/}. After coding, the inter-coder disagreement was at 5\%, which we resolved by majority with a third coder. In this random sample, we found 54 meta-comments (5.4\%) of which only five addressed the community-moderator. The second column of Table  \ref{tab:labeled_comment_counts} summarizes the label distribution for this random sample. We interviewed the coders to deduce machine learning features from their observations.
 
\begin{table}[tbh]
\caption{The number of each label in the random sample, the SPON training set, and the OMP training set.}
\label{tab:labeled_comment_counts}
\centering
\footnotesize
\begin{tabular}{l|r|r|r}
\toprule
 & & \multicolumn{2}{c}{\bf{Training Sets}} \\
\textbf{Labels} & \textbf{Random Sample} & \textbf{SPON} & \textbf{OMP} \\ 
\midrule
\rowcolor{Gray}     Media       &  25 &  404 &  566 \\ 
                    Journalist  &  33 &  426 &  198 \\ 
\rowcolor{Gray}     Moderator   &   5 &  323 &  421 \\ 

\midrule
                    Meta            &  54 &   982  & 1,301 \\
\rowcolor{Gray}     Non-Meta        & 946 & 1,127 & 4,737 \\ 
\midrule
\textbf{Total}          & \textbf{1,000} & \textbf{2,109} & \textbf{6,038} \\ 
\bottomrule
\bottomrule
\end{tabular}
\end{table} 

\section{Feature Deduction}
\label{sec:freatures-eng}
We describe the training of word and comment embeddings as well as the machine learning features, which we derived from the insights of our qualitative content analysis.


\subsection{Training Word and Comment Embeddings}
Word embeddings are a geometric way of capturing the meaning of a word by using low-dimensional vectors  
\cite{rumelhart1988learning}. Their main advantage is that the vector representation of similar words are situated close in vector space. We used word2vec \cite{mikolov2013efficient} to obtain a distributed vector representation for German user comments. As an input word2vec requires a text corpus as large as possible to produce low-dimensional vectors as an output. Besides word2vec, paragraph2vec (or doc2vec) \cite{le2014distributed} produces document embeddings from comments or articles. We used the Python library gensim \cite{rehurek_lrec} to generate the embeddings.

We preprocessed the comments in four steps: (1) concatenated each comment's title with its text, (2) removed stop words, (3) removed punctuation, and (4) converted the text to lower case. We noted that for word2vec, using more than 300 dimensions or a window size of more than 5 unnecessarily increases the training time while not improving the precision of the vector representation \cite{pennington2014glove}. 

We used three different word embedding models for the end-to-end learning approach. Table \ref{tab:word2vec_models} compares our generated SPON model with two other models: the OMP model according to Schabus et al.~\cite{Schabus2017}, and the GermanWord model that M\"uller \cite{mueller2015} trained on German Wikipedia and news articles.

We used the SPON user comments to train both (1) word embeddings with word2vec and (2) comment embeddings with doc2vec. We used the word embeddings to enrich a set of keywords and to pre-fill the embedding layer of a neural network (transfer learning) \cite{michalski1983theory}. We used the comment embeddings to extract semantic features. To enable replication, our models are publicly available on our project website.

\begin{table}[tbh]
\caption{A comparison of the training parameters between the three different word2vec models we used.}
\label{tab:word2vec_models}
\renewcommand{\arraystretch}{1.1}
\centering
\footnotesize

\begin{tabular}{l|r|r|r}
\toprule
\textbf{Model} & \textbf{SPON} & \textbf{OMP} & \textbf{GermanWord} \\ 
\midrule
\rowcolor{Gray} Number of dimensions    &         300 &        300 &         300  \\ 
                Vocab size              &     212,630 &    129,070 &     608,130  \\ 
\rowcolor{Gray} Corpus size in words    & 462,269,114 & 31,489,845 & 651,219,519  \\
                Min count               &          50 &          5 &           5  \\ 
\rowcolor{Gray} Window size             &           5 &          5 &           5  \\
                Training epochs         &           5 &         10 &          10  \\ 
\rowcolor{Gray} Training method         &        CBOW &       CBOW &   Skip-gram  \\

\bottomrule
\end{tabular}
\end{table}

\subsection{Machine Learning Features}
We categorize all used machine learning features for our dataset into three groups: text features, semantic features, and metadata. We indicate the features specific to the SPON dataset with [S].

\subsection*{Text Features}
In the following, we list the text features we identified based on the coders' insights from the qualitative content analysis and related work discussing the criteria media organizations consider when identifying high-quality comments \cite{jorgensen_understanding_2002, mcenroy_where_2013, reader_air_2007}. Diakopoulos \cite{diakopoulos2015editor, diakopoulos2015picking} categorized these criteria into twelve human-centric categories, including  emotionality, readability, thoughtfulness, brevity, and novelty. 

\begin{itemize}
  \item[--] \textbf{Regular expression pattern}: We identified a set of keywords based on word embeddings, which are likely to be used in meta-comments. We followed a two-step approach: (1) manual keyword collection and (2) keyword enrichment with word embeddings. We used the SPON word embeddings and fine-tuned the keywords for the SPON dataset. We started by manually collecting an initial set of keywords with communication researchers. Given the vector representations of the words in comment texts, we enriched the manually collected keywords by finding the most similar words (see Table \ref{tab:word2vec_word_examples}). This shows how word embeddings can capture further words with a similar meaning and common misspellings. We created a regular expression (regex) based on the keywords to match words independently of the grammatical gender. We iteratively searched for user comments that match the regex pattern, assessed the matching comments, and adjusted the regex pattern to minimize unintended matches. We list the translated set of keywords for each meta-addressee class: (media) media, spon, spiegel, spiegelonline, editing, reporting, magazine; (journalist) article, journalism, contribution, author, writer, editor, penpusher, columnist, expert, reporter, spiegel editor, populist, last names of the SPON authors; (community-moderator) censorship, censored, moderation, moderator, admin, sysop.
  
  
      
      
      
    
    \begin{table}[tbh]
    \caption{Examples of similar words within the distributed vector space for the last name of the journalist ``Mr. Fleischhauer'' and the word ``autor'' (author).}
    \label{tab:word2vec_word_examples}
    \renewcommand{\arraystretch}{1.0}
    \centering
    \small
    \begin{tabular}{l|r || l|r}
    	\toprule
    	\bf{Word} & \bf{Similarity} & \bf{Word} & \bf{Similarity}\\
    	\midrule
        \rowcolor{Gray} \textbf{fleischhauer} & 1.00         & \textbf{autor}            & 1.00       \\
                              fleischauer & 0.91         &             author & 0.86       \\
        \rowcolor{Gray}           augstein & 0.88         &          verfasser & 0.85       \\
                                      lobo & 0.82         &       spiegelautor & 0.80       \\
        \rowcolor{Gray}               diez & 0.80         &   artikelschreiber & 0.80       \\
                                  matussek & 0.77         &          sponautor & 0.80       \\
        \rowcolor{Gray}      fleischhauers & 0.77         &            autorin & 0.80       \\
                                  kuzmany & 0.76         &        verfasserin & 0.72       \\
        \rowcolor{Gray}        fleichhauer & 0.76         &      schreiberling & 0.71       \\
                                 m\"unchau & 0.76         &          rezensent & 0.70       \\
        \rowcolor{Gray}              dietz & 0.75         &          schreiber & 0.69       \\
                                    nelles & 0.73         &   spiegelredakteur & 0.69       \\
        \rowcolor{Gray}             broder & 0.73         &        kommentator & 0.68       \\
                                  mattusek & 0.71         &          kolumnist & 0.68       \\
        \rowcolor{Gray}          mattussek & 0.71         &       artikelautor & 0.68       \\
                                     kaden & 0.70         &          redakteur & 0.65       \\
        \rowcolor{Gray}          neubacher & 0.70         &      sponredakteur & 0.64       \\
                                    fricke & 0.70         &   artikelverfasser & 0.64       \\
        \rowcolor{Gray}            rickens & 0.69         &             forist & 0.63       \\
    	
    	\bottomrule
    	
    \end{tabular}
    \end{table}

    
    

  \item[--] \textbf{Tf-idf}: 
  The tf-idf score of a word reveals the importance of this word in a user comment. It assigns words a greater weight proportionally to the occurrence frequency but reduces the significance of a word that frequently occurs in many documents as stop words. We used the tf-idf representation of the comment with unigrams and bigrams without stop words.

  \item[--] \textbf{Count of ``Sie'' occurrences}: In the German language, the formal address of ``you'' to an unknown person is ``Sie'' and is written with a capital ``S'' even if it is situated within a sentence. We count the occurrences of this address within the sentence to separate it from the similar third-person pronoun ``sie''. For the identification of each occurrence, we used the regular expression pattern ``\verb_[^\.!?]\s+Sie_''. We assumed that it is an indicator of a reference to the article's author. However, our coders observed that this formal address often refers to other users. For this, commenters also use the ``@'' notation to indicate a reference.

  \item[--] \textbf{Number of questions}: Questions in comment texts might address the media company, authors, or community-moderators. Our coders mentioned typical user questions as ``\textit{Why has my comment been blocked?}''. Therefore, we identified and counted the number of questions, contained in a comment. 
  
  \item[--] \textbf{Length}: We added together the number of characters in the comment title and text. We assumed that meta-comments might differ in their length from other user comments as previous work has also identified brevity as a quality indicator.
  
  \item[--] \textbf{Average word length}: We used the average number of characters per word as a simple measure of text complexity. Users might put more effort in the wording of a user comment and choose more sophisticated and longer words on average in meta-comments.
  
  \item[--] \textbf{Number of capital letters}: We count the number of capital letters. Users often use capital letters to indicate ``yelling'' in user comments. We assumed that these comments are more likely to complain about meta-addressees. Besides, users also write the names of the media companies in capital letters such as ``SPIEGEL'' or ``DER STANDARD''.
  
  \item[--] \textbf{Sentiment score}: We used the sentiment score \cite{sentiment_textblob} of the comment title and text, assuming that a high polarity score is an indicator of media-critical statements \cite{craft2016reader}.
\end{itemize}

\subsection*{Semantic Features}
We used two different semantic features, derived from comment embeddings:
\begin{itemize}
  \item[--] \textbf{Document vector}: From paragraph2vec, we obtained a 300-dimensional dense vector representation for each comment in a distributed vector space in which semantically similar comments have a high cosine similarity. We used each dimension of this vector as a feature. As we generated the comment embeddings based on the SPON user comments, the model infers a vector representation for the OMP comments as we did not use them for training.
  
  \item[--] \textbf{Vector Space Distance}: We utilized the comment embeddings to determine a representative average vector (class vector) for each comment class. We used the cosine distance and the most similar class vector as a feature. We formally describe the semantic distance feature. Let $A$ be the set of all comments and $\mathcal{C}$ the set of all comment classes. Further, let $W: A\rightarrow \mathbb R^{300}$ be the comment embedding function that yields a vector representation for a comment. Then, for each class $X \in \mathcal{C}$ we define a class vector $\overline{X}$, which is an average vector as follows: 
$$\overline{X}=\dfrac{1}{|X|} \sum_{c\in X} W(c)$$
As a feature for a comment $c \in A$, we used the cosine distance function $d$ to determine the distance $d(W(c), \overline{X})$ for each $X \in \mathcal{C}$. Additionally, we identified the class $X$ to which the class vector $\overline{X}$ has the minimal distance $\min\limits_{X\in \mathcal{C}} d(W(c), \overline{X})$ and added it one-hot encoded as a feature.
\end{itemize}

\subsection*{Metadata}
The metadata is the set of additional properties of a user comment. We obtained more additional metadata for SPON user comments. We extracted the following features from the metadata:
\begin{itemize}
  \item[--] \textbf{Comment number [S]}: The forum lists the user comments in ascending order of time, assigning each comment an consecutive number. This number is the position of the comment in the list. We added the comment position as a feature, as first user comments might be more likely to identify errors in the article.
  \item[--] \textbf{Department [S]}: The SPON page is structured into twelve departments. As users post their comments to an article, we used the department of the article as a feature.
  \item[--] \textbf{Quote contained [S]}: Users can reply to comments from other users. With this function, users can quote a previous user's comment text. We assumed that users instead address another user than a meta-addressee when they refer to other comments. This assumption corresponded with our coders' impressions.
  \item[--] \textbf{Time}: We further extracted the time stamp precisely to the minute of each comment. We add both the day of the week and the hour of the day as features.
\end{itemize}

\section{Classifier Experimentation and Optimization}
\label{sec:cla-exp-opt}
We used a supervised machine learning approach for the user comment classification. The classifier derives a classification model from these labeled training sets to classify unseen user comments. The training set contains comments with the label meta-comment (with meta-addresses) or non-meta-comment. Our approach uses four binary classifiers in two steps: (1) a binary classifier for meta-comment / non-meta-comment and (2) three binary classifiers to classify each meta-addressee class. For the second step, we used the classification strategy one-vs-all \cite[p. 182,338]{bishop2006pattern}, which trains a binary classifier per class.

\subsection{Training Set Creation}
For the SPON training set, we collected coded comments for each meta-addressee class. Due to the small share of meta-comments, random sampling was not feasible for gathering enough comments per meta-addressee class. For sampling a user comment set with a higher share of meta-comments for annotation, we used (1) regular expressions and (2) cosine similarity between keywords and user comments in the vector space of the comment embeddings. We calculated the average vector of the keywords for each meta-addressee class and labeled the 100 most similar comments to each average vector. With this approach, we captured a heterogeneous set of user comments, for which manual labeling was feasible. We used the non-meta-comments of the random sample as well as the non-meta-comments of the sampling described above.

For the OMP dataset, we followed the same coding procedure to identify the meta-addressees for the 1301 feedback comments. Table \ref{tab:labeled_comment_counts} shows the distribution of meta-comments and meta-addressee comments for our SPON and OMP training sets. The latter contains 240 comments, which we were unable to assign to a meta-addressee class.

\subsection{Classification Approaches}
We compare the user comment classification results between a traditional machine learning approach and an end-to-end learning approach based on a neural network model. While the traditional classification approach requires a data representation based on hand-crafted features, neural networks can handle raw text as an input and learn high-level feature representations automatically \cite{Goodfellow-et-al-2016}. They have been applied with remarkable results in different classification tasks as object detection in images, machine translation, sentiment analysis, and text classification tasks \cite{collobert2011natural}.

\begin{figure}[hbt]
  \includegraphics[width=\textwidth]{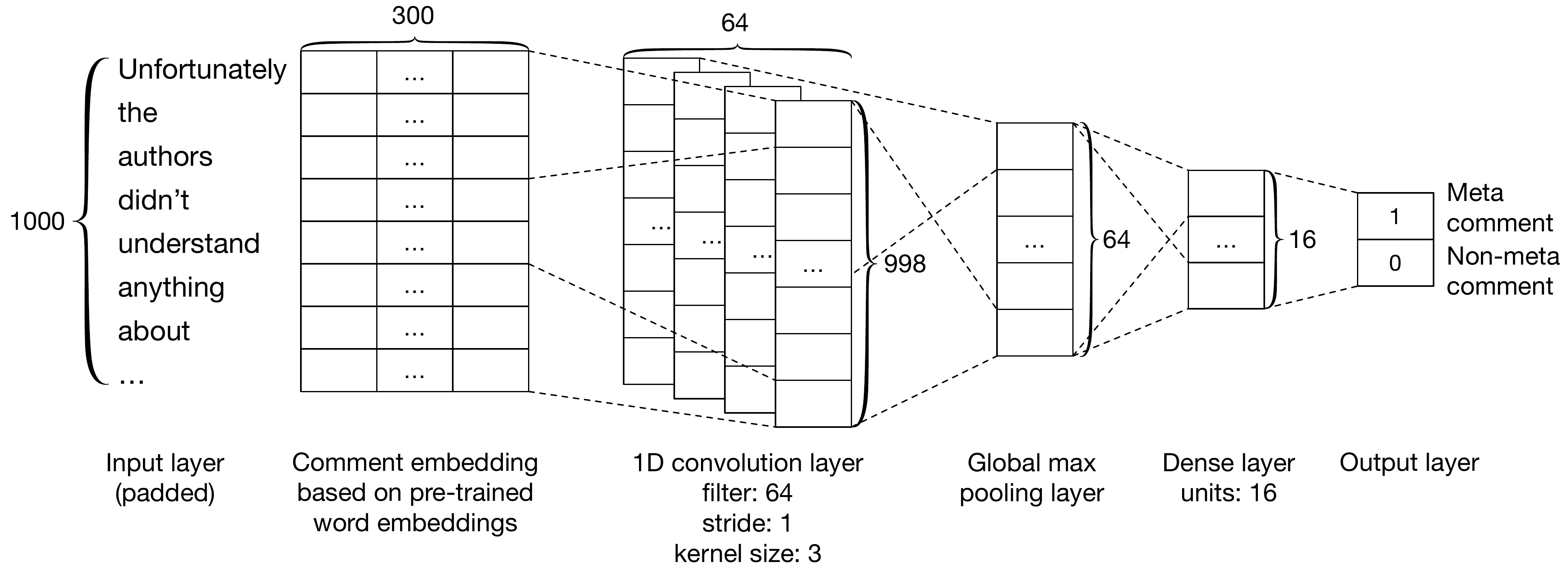}
  \caption{Neural network architecture with optimized hyperparameters for the user comment classification.}
  \label{fig:cnn_model}
\end{figure}

Convolutional neural networks have mainly been used for image classification tasks, but researchers have also started using them to solve natural language processing tasks \cite{kim2014convolutional}. Given the small training set for an end-to-end approach, we used a shallow neural network model and experimented with different numbers of epochs to prevent the model from overfitting. We padded the input comment text to a maximum length of 1,000 words. As shown in Figure \ref{fig:cnn_model}, after the input layer our network consists of an embedding layer, a 1D convolution layer, a 1D global max pooling layer, a dense layer, and a concluding output layer with a softmax activation. For the other layers, we used the tanh activation function. 
We applied transfer learning  \cite{michalski1983theory} by pre-initializing the embedding layer of the model with three different word2vec models, which we compared in Table \ref{tab:word2vec_models}. While training the model, we froze the weights of the embedding layer.

Due to the small size of our training set, we conducted a stratified 10-fold cross-validation on the training set to acquire reliable results. For assessing the classification results, we report on precision, recall (to compare our results with state-of-the-art results) and the $F_\beta$ measure (to overvalue precision over recall). 
For the experiments, we used the Python libraries scikit-learn \cite{scikit-learn} for the traditional approach and Keras \cite{chollet2015keras} for the end-to-end approach.

\subsection{Hyperparameter Optimization}
To answer RQ1, we performed a grid search to optimize the hyperparameters for both classification approaches. A grid search performs an exhaustive search over specified hyperparameter values for a classifier. We evaluated each parameter combination with a stratified three-fold cross-validation to reduce the computational complexity. To enable replication, the relevant source code, containing the parameter grids for both approaches are publicly available on our project website.

We value precision over recall to minimize type I errors (false positives) for the end user so that the comment analyst has to read a minimal number of wrongly-classified meta-comments. The classifier might not catch all meta-comments, but on the other hand, we minimize the time spent by the analyst reading irrelevant comments. We used the $F_\beta$ score as the scoring method for the grid search. It is the weighted harmonic mean of precision and recall \cite[pp.327-328]{yates2011modern}. 
We specify $\beta = 0.5$ to overvalue the precision score in our evaluation metric. We compare the accuracy of five different classifiers.

\begin{itemize}
    \item[--] \textbf{Support Vector Machine} (SVM) is known to be one of the best text classifiers found in the literature \cite{ben2010user}. 
    
    \item[--] \textbf{Decision Tree} learning assumes that all features have finite discrete domains and that there is a single target feature representing the classification (i.e., the tree leaves) \cite{torgo2016data}. 
    
    \item[--] \textbf{Random Forest} \cite{breiman2001random} is a combination of decision tree classifiers on sub-samples and controls over-fitting. 
    
    \item[--] The meta-classifier \textbf{AdaBoost} \cite{freund1995desicion} initially fits a classifier on the original dataset and then fits additional copies of the classifier, adjusting the weights for wrongly classified samples. 
    
    \item[--] \textbf{KNeighbors} does not construct a general model, but stores the training data and the classification for a point, which is derived from a majority vote of all nearest neighbors \cite{cunningham2007k}. 
\end{itemize}

We additionally varied the number of the most significant features for each classifier to 10, 50, and ``all features''. We conducted multiple grid search runs and added more fine-grained values into the parameter ranges to find the parameters for the best results.
 
The performance of neural networks is dependent on their architecture as well as the right hyperparameter selection. To optimize the neural network architecture, we also performed a grid search over the combined dataset and evaluated each configuration with a stratified three-fold cross-validation. We achieved the best results with the neural network architecture depicted in Figure \ref{fig:cnn_model}, trained with a batch size of 32 for 5 epochs. 

\subsection{User Comment Classification}
The grid search results showed that SVM with a linear kernel using all machine learning features achieves the best results for the SPON dataset, the OMP dataset, and the combined dataset. For the SPON and the combined dataset, the penalty parameter $C=0.5$ achieves the best $F_{0.5}$ values, for the OMP dataset $C=1.0$. The results in Table \ref{tab:classifier_comparison_results} show that the traditional classification approach outperforms the end-to-end learning approach for the SPON dataset ($F_{0.5}=0.91$) and the combined dataset ($F_{0.5}=0.87$). The end-to-end approach outperforms the traditional approach on the OMP dataset ($F_{0.5}=0.85$) pre-initialized with either the SPON word embedding model or the OMP model. However, the performance difference between the traditional and the end-to-end approach is negligible ($\Delta F_{0.5}\leq0.05$).

The results show a higher $F_{0.5}$ score if we use pre-trained word embeddings based on user comments rather than embeddings based on the Wikipedia and news corpora. It is also striking that we achieve the same $F_{0.5}$ scores with both the SPON and OMP embeddings. Schabus et al.~\cite{Schabus2017} have also compared different classification approaches on the Feedback category of the OMP dataset where they achieved a best precision of 0.75, a recall of 0.71, and an F1-score of 0.63. All of our classification results outperformed their state-of-the-art results by up to 11\% for precision and 12\% for recall.

\begin{table*}[tbh]
\caption{User comment classification (meta / non-meta) results of a stratified 10-fold cross validation for three different training set compositions.}
\label{tab:classifier_comparison_results}
\renewcommand{\arraystretch}{1.0}
\begin{center}
\resizebox{\linewidth}{!}{%

\begin{tabular}{  l | r r r | r  r  r | r  r  r }
	\toprule
	\bf{User Comment Classification Approach} & \multicolumn{3}{c|}{\bf{SPIEGEL Online}} & \multicolumn{3}{c|}{\bf{One Million Posts}} &  \multicolumn{3}{c}{\bf{Combined Dataset}} \\\cline{2-10}
	& \footnotesize Precision &  \footnotesize Recall & \footnotesize $F_{0.5}$ & \footnotesize Precision &  \footnotesize Recall & \footnotesize $F_{0.5}$ & \footnotesize Precision &  \footnotesize Recall & \footnotesize $F_{0.5}$  \\ \midrule
	
	\rowcolor{Gray}
	Traditional (with manual features)
	& 0.91 &   0.91 &   \textbf{0.91}    
	& 0.83 &   0.81 &   0.82             
	& 0.88 &   0.82 &   \textbf{0.87}\\  
		
	End-to-End (with SPON embeddings)
	& 0.86 &   0.87 &   0.86             
	& 0.86 &   0.81 &   \textbf{0.85}    
	& 0.85 &   0.83 &   0.85\\           
	
	\rowcolor{Gray}
	End-to-End (with One Million embeddings)
	& 0.84 &   0.82 &   0.84             
	& 0.85 &   0.83 &   \textbf{0.85}    
	& 0.84 &   0.81 &   0.84\\           
	
	End-to-End (with GermanWord embeddings)
	& 0.73 &   0.77 &   0.73    
	& 0.77 &   0.73 &   0.76    
	& 0.73 &   0.74 &   0.73\\  
	
	\bottomrule
	
\end{tabular}}
\end{center}
\end{table*}

\subsection{Meta-Comment Classification}
For the second step, we classified meta-comments with regards to their meta-addressees for the SPON and OMP datasets. We used SVM with a linear kernel and the penalty parameter set to $C=0.5$ as it achieved the best results for the user comment classification. Table \ref{tab:feature_group_comparison} shows the results for both datasets as well as the classification results using different feature groups, which we describe later. 
The SPON dataset classification achieved high scores with $F_{0.5}\geq0.84$ for all meta-addressee classes. The $F_{0.5}$ scores for the SPON dataset are higher than the OMP dataset. For the Media and the Moderator class, the differences between the datasets are minor with $\Delta F_{0.5}\leq0.06$.

\begin{table*}[tbh]
\caption{User comment and meta-comment classification results of a stratified 10-fold cross-validation for both training sets, using an SVM classifier with different feature groups.}
\label{tab:feature_group_comparison}
\renewcommand{\arraystretch}{1.15}
\begin{center}
\resizebox{\linewidth}{!}{%

\begin{tabular}{ c | l | r r r | r  r  r | r  r  r |  r  r  r  }
	\toprule
	& \bf{Feature Combination} & \multicolumn{3}{c|}{\bf{Meta}} & \multicolumn{3}{c|}{\bf{Media}} &  \multicolumn{3}{c|}{\bf{Journalist}} & \multicolumn{3}{c }{\bf{Moderator}}  \\\cline{3-14}
	& & \footnotesize Precision &  \footnotesize Recall & \footnotesize $F_{0.5}$ & \footnotesize Precision &  \footnotesize Recall & \footnotesize $F_{0.5}$ & \footnotesize Precision &  \footnotesize Recall & \footnotesize $F_{0.5}$  & \footnotesize Precision &  \footnotesize Recall & \footnotesize $F_{0.5}$ \\ \midrule
	
	\rowcolor{Gray} \cellcolor{white!30}
	& All 
	& 0.91 &   0.91 &   \textbf{0.91}          
	& 0.85 &   0.81 &   \textbf{0.84}          
	& 0.88 &   0.76 &   \textbf{0.86}          
	& 0.84 &   0.87 &   \textbf{0.84}\\        
		
	& Without regex patterns    
	& 0.82 &   0.80 &   0.82    
	& 0.80 &   0.63 &   0.76    
	& 0.73 &   0.55 &   0.68    
	& 0.84 &   0.68 &   0.80\\  

    \rowcolor{Gray}	\cellcolor{white!30}
	& Only regex patterns 
	& 0.90 &   0.93 &   \textbf{0.91}          
	& 0.84 &   0.85 &   \textbf{0.84}          
	& 0.89 &   0.69 &   0.84                   
	& 0.82 &   0.86 &   0.82\\                 
	
    \multirow{-4}{*}{\rotatebox[origin=c]{90}{\textbf{SPIEGEL}}}
	& Only semantic features 
	& 0.77 &     0.71 &     0.76   
	& 0.76 &     0.36 &     0.62   
	& 0.68 &     0.42 &     0.61   
	& 0.75 &     0.34 &     0.60\\ 
	
	\midrule
	
	\rowcolor{Gray} \cellcolor{white!30}
	& All 
	& 0.85 &   0.79 &   \textbf{0.84}   
	& 0.79 &   0.82 &   \textbf{0.79}   
	& 0.78 &   0.39 &   \textbf{0.65}   
	& 0.81 &   0.67 &   \textbf{0.78}\\ 
		
	& Without regex patterns            
	& 0.81 &   0.80 &   0.81            
	& 0.76 &   0.83 &   0.77            
	& 0.79 &   0.38 &   \textbf{0.65}   
	& 0.82 &   0.68 &   \textbf{0.78}\\ 

    \rowcolor{Gray} \cellcolor{white!30}
	& Only regex patterns      
	& 0.88 &   0.44 &   0.73   
	& 0.74 &   0.53 &   0.69   
	& 0.89 &   0.09 &   0.31   
	& 0.85 &   0.07 &   0.25\\ 
	
    \multirow{-4}{*}{\rotatebox[origin=c]{90}{\textbf{One Million}}}
	& Only semantic features 
	& 0.73 &     0.62 &     0.70   
	& 0.63 &     0.80 &     0.66   
	& 0.74 &     0.17 &     0.45   
	& 0.73 &     0.47 &     0.66\\ 

	\bottomrule
	
\end{tabular}}
\end{center}
\end{table*}

We also performed a \textbf{cross-dataset} classification. We trained the binary classifiers with the SPON dataset (training set) and classified the labeled user comments of the OMP dataset (test set) and vice versa. Table \ref{tab:cross_data_source_results} shows the results. The $F_{0.5}$ scores are higher for all classes when trained on the OMP dataset and applied to the SPON dataset. The recall values were low for all classes ($<0.4$) when using the SPON training set.

\begin{table*}[tbh]
\caption{Cross-dataset classification results of an SVM classifier trained with the SPIEGEL Online data and applied on the OMP dataset and vice versa.}
\label{tab:cross_data_source_results}
\renewcommand{\arraystretch}{1.0}
\begin{center}
\resizebox{\linewidth}{!}{%

\begin{tabular}{ l | l | r r r | r  r  r | r  r  r |  r  r  r  }
	\toprule
	\bf{Training Set} & \bf{Test Set} & \multicolumn{3}{c|}{\bf{Meta}} & \multicolumn{3}{c|}{\bf{Media}} &  \multicolumn{3}{c|}{\bf{Journalist}} & \multicolumn{3}{c}{\bf{Moderator}}  \\\cline{3-14}
	& & \footnotesize Precision &  \footnotesize Recall & \footnotesize $F_{0.5}$ & \footnotesize Precision &  \footnotesize Recall & \footnotesize $F_{0.5}$ & \footnotesize Precision &  \footnotesize Recall & \footnotesize $F_{0.5}$  & \footnotesize Precision &  \footnotesize Recall & \footnotesize $F_{0.5}$ \\ \midrule
	
	\rowcolor{Gray}
	SPIEGEL Online & One Million Posts
	& 0.90 &   0.38 &   0.71          
	& 0.82 &   0.22 &   0.53          
	& 0.38 &   0.33 &   0.37          
	& 0.59 &   0.34 &   0.51\\        
		
	One Million Posts & SPIEGEL Online
	& 0.89 &   0.71 &   \textbf{0.85}    
	& 0.63 &   0.88 &   \textbf{0.67}    
	& 0.82 &   0.60 &   \textbf{0.76}    
	& 0.87 &   0.75 &   \textbf{0.84}\\  
	
	\bottomrule
	
\end{tabular}}
\end{center}
\end{table*}

We tested the accuracy of the meta-comment classifier on unseen comments by classifying a random sample of 100,000 SPON comments regarding the three meta-addressee classes. The classifier assigned a label to a comment when the confidence score is greater than 0.8. In a comment analytics tool, this could be a user-adjustable parameter. Instead of ranking the labeled comments according to the confidence score we randomly selected 300 meta-comments (100 per meta-addressee). Following the coding guide (Section \ref{sec:qual_cont_anal}), the same coders manually checked if the classification was correct. This application would be similar to a desirable use case for comment analysts \cite{loosen2017exploratory}. We achieved the following accuracy: 0.94 (Media), 0.64 (Journalist), and 0.67 (Moderator).

\subsection{Feature Significance}
To answer RQ2, we calculated the analysis of variance (ANOVA) F-value for each single machine learning feature and sorted them accordingly as shown in Table \ref{tab:feature_significance_per_meta_addressee}.  
For the SPON dataset, the most significant feature for the meta-comment identification is the meta property ``department\_career''. In our training set, we found only 35 meta-comments posted on the career department. The results show that our extended regular expression set is a significant feature of the SPIEGEL dataset and achieves an $F_{0.5}$ score of 91\% for the meta-comment class as well as scores between 82\% and 84\% for the meta-addressee classes. The regex patterns for each meta-addressee class are the most important features respectively. Other essential features are the tf-idf scores of uni-grams. Not a single tf-idf bigram is in the list.

\begin{table*}[tbh]
\caption{Top ten single features for classifying user and meta-comments according to their ANOVA F-value.}
\label{tab:feature_significance_per_meta_addressee}
\renewcommand{\arraystretch}{1.2}
\centering
\resizebox{\textwidth}{!}{%
\begin{tabular}{c|lr|lr|lr|lr}

\toprule

& \multicolumn{2}{c|}{\textbf{Meta}} & \multicolumn{2}{c|}{\textbf{Media}} & \multicolumn{2}{c|}{\textbf{Journalist}} & \multicolumn{2}{c}{\textbf{Moderator}} \\

\midrule

\rowcolor{Gray} \cellcolor{white!30} & department\_carreer            &  437 & regex\_media\_matches          & 390 &  regex\_journalist\_matches      & 167 & regex\_moderator\_matches     &  680 \\
                                     & regex\_journalist\_matches     &  328 & keyword\_\textbf{spon}         & 181 &  regex\_moderator\_matches       & 162 & keyword\_\textbf{sysop}       &  206 \\
\rowcolor{Gray} \cellcolor{white!30} & regex\_media\_matches          &  206 & tfidf\_\textbf{spiegel}        & 110 &  tfidf\_\textbf{herr}            &  58 & tfidf\_\textbf{zensiert}      &   95 \\
                                     & regex\_moderator\_matches      &  138 & keyword\_\textbf{spiegel}      &  84 &  keyword\_\textbf{sysop}         &  48 & tfidf\_\textbf{sysop}         &   80 \\
\rowcolor{Gray} \cellcolor{white!30} & keyword\_\textbf{spon}         &  123 & keyword\_\textbf{redaktion}    &  66 &  keyword\_\textbf{zensiert}      &  40 & keyword\_\textbf{zensiert}    &   77 \\
                                     & tfidf\_\textbf{spiegel}        &   84 & tfidf\_\textbf{redaktion}      &  53 &  tfidf\_\textbf{zensiert}        &  40 & tfidf\_\textbf{beitrag}       &   71 \\
\rowcolor{Gray} \cellcolor{white!30} & keyword\_\textbf{artikel}      &   83 & tfidf\_\textbf{medien}         &  51 &  department\_carreer             &  35 & keyword\_\textbf{zensur}      &   71 \\
                                     & text\_capitalletters           &   80 & tfidf\_\textbf{spon}           &  50 &  keyword\_\textbf{spon}          &  32 & tfidf\_\textbf{beitr\"age}    &   60 \\
\rowcolor{Gray} \cellcolor{white!30} & tfidf\_\textbf{artikel}        &   78 & keyword\_\textbf{sysop}        &  48 &  regex\_media\_matches           &  32 & keyword\_moderation           &   59 \\

\multirow{-10}{*}{\rotatebox[origin=c]{90}{\textbf{SPIEGEL Online}}}
                                     & keyword\_\textbf{spiegel}      &   78 & regex\_moderator\_matches      &  43 &  keyword\_\textbf{zensur}        &  31 & keyword\_\textbf{beitrag}     &   59 \\

\midrule

\rowcolor{Gray} \cellcolor{white!30} &  tfidf\_\textbf{standard}        &  302 &  tfidf\_\textbf{standard}                & 181 &  tfidf\_\textbf{herr}              & 173 & semantic\_min\_dist\_moderator     & 174 \\
                                     &  regex\_journalist\_matches      &  257 &  regex\_media\_matches                   &  67 &  tfidf\_\textbf{rauscher}          & 147 & tfidf\_\textbf{postings}           &  91 \\
\rowcolor{Gray} \cellcolor{white!30} &  semantic\_min\_dist\_non-meta   &  212 &  semantic\_min\_dist\_moderator          &  54 &  semantic\_min\_dist\_journalist   &  82 & tfidf\_\textbf{gel\"oscht}         &  67 \\
                                     &  semantic\_min\_dist\_meta       &  212 &  tfidf\_\textbf{artikel}                 &  51 &  tfidf\_\textbf{herr rauscher}     &  77 & tfidf\_\textbf{posting}            &  53 \\
\rowcolor{Gray} \cellcolor{white!30} &  keyword\_\textbf{artikel}       &  207 &  text\_avgwordlength                     &  48 &  tfidf\_\textbf{frau}              &  76 & tfidf\_\textbf{artikel}            &  52 \\
                                     &  tfidf\_\textbf{artikel}         &  194 &  tfidf\_\textbf{postings}                &  47 &  text\_num\_sie                    &  63 & semantic\_sem\_16                  &  49 \\
\rowcolor{Gray} \cellcolor{white!30} &  keyword\_\textbf{redaktion}     &   88 &  semantic\_min\_dist\_media              &  45 &  semantic\_sem\_236                &  46 & tfidf\_\textbf{posts}              &  48 \\
                                     &  regex\_media\_matches           &   81 &  keyword\_contained\_\textbf{artikel}    &  42 &  tfidf\_\textbf{standard}          &  42 & tfidf\_\textbf{standard}           &  48 \\
\rowcolor{Gray} \cellcolor{white!30} &  tfidf\_\textbf{redaktion}       &   79 &  tfidf\_\textbf{gel\"oscht}              &  41 &  semantic\_sem\_158                &  40 & regex\_journalist\_matches         &  47 \\

\multirow{-10}{*}{\rotatebox[origin=c]{90}{\textbf{One Million Posts}}}
                                     &  text\_avgwordlength             &   65 &  keyword\_contained\_\textbf{redaktion}  &  40 &  keyword\_contained\_\textbf{Rau}  &  36 & semantic\_min\_dist\_media         &  47 \\

\bottomrule

\end{tabular}%
}%
\end{table*}

In the OMP dataset, the minimal semantic distance is among the top ten significant features for all classes. ``Herr Rauscher'' (Mr. Rauscher) is a journalist for the Austrian news site. The tf-idf bigram score for ``herr rauscher'' is significant for the Journalist class. Also, the regex sets for Journalist and Media are among the top features. The text feature \textit{average word length} appears in the list of the Meta and Media class. The text feature \textit{occurrence of ``Sie''} appears in the Journalist class.

For both datasets, we can see that the names of the media company are significant features: ``spon'', ``spiegel'', and ``standard''. We assume that the bigram ``der standard'' is not in the list because we removed stop words, which also contain the German article ``der'' (the). The words ``artikel'' (article), ``redaktion'' (editing), and ``herr'' (mr.) are significant features for both datasets.

In Table \ref{tab:feature_group_comparison} we compare four different feature groups using an SVM classifier as the baseline with a linear kernel and the penalty parameter $C=0.5$. We also performed a stratified 10-fold cross-validation to acquire the precision, recall, and $F_{0.5}$ score for the classification.

For the SPON dataset, the regex-based features achieve high results. The improvement of further features is minor. By adding the remaining features, the $F_{0.5}$ score increased up to 2\% (for Moderator). For the Journalist class, the regex patterns are an essential feature and the $F_{0.5}$ score reduced drastically when they were removed. Further, additional features do not improve the $F_{0.5}$ score. Semantic features by themselves achieve an $F_{0.5}$ score of up to 76\% on SPON meta-comments.

In the OMP dataset, the regex features are not relevant for the classes Journalist and Moderator and barely relevant for Meta and Media with $\Delta F_{0.5}\leq0.03$. The Journalist class achieves the lowest $F_{0.5}$ score of 0.65. The Media and Moderator class achieve a similar $F_{0.5}$ score of 0.79 and 0.78.

\section{Qualitative Insights into Classified Meta-Comments}
\label{sec:qualitative}
To answer RQ3, we describe examples from the content of correctly classified meta-comments (true positives) from both datasets, a qualitative method inspired by Kurtanovi{\'c} and Maalej \cite{kurtanovic2017mining}. The purpose of this qualitative analysis is to understand the content and the potential usefulness of meta-comments. We classified meta-comments for each meta-addressee class and dataset and identified different information types. 
We translate the user comments into English.

\subsection{Comments Addressing the Media}
The meta-comments addressing the media criticize the prioritization of the media company. These users \textbf{demand justification} for the attention the authors pay to a particular topic (e.g.~\#1,\#2), \textbf{report an error} in the article text (e.g.~\#3), and \textbf{praise} the media coverage (e.g.~\#4):

\#1 SPON: \textit{``[...], but it gets a whole article in the Spiegel. Please, someone explain this over-dramatization! It shows, however, that the drug policy and the anti-drug laws are lacking in goals and are, therefore, practically nonsense, but both have a lot of support from the press (Spiegel?). [...]''}

\#2 SPON: \textit{``[...] it's just disgusting, how journalists in Germany keep themselves busy and can seriously make a big thing out of this farce. Words fail me, that something like this does not appear as a 3-line message in the furthest corner of a tabloid newspaper, [...]''}

\#3 OMP: \textit{`` ``They complete reconnaissance aircrafts.'' How does such an article come about? Is this proofread or will you press Enter after the last word and go to the coffee machine?''}

\#4 OMP: \textit{``Thanks, mka for the background. Most media have always only reported on the prayer room, and nebulously mentioned that the day before firefighters and a police officer had been injured, but neither how, where, in what context. Like this article, I want journalism.''}

\subsection{Comments Addressing the Journalist}
The listed classified meta-comments addressing the journalist contain \textbf{praise} (\#5), \textbf{recommendations} for other readers (\#5), \textbf{further questions} (\#6), \textbf{missing information} (\#7), \textbf{critiques} (\#6,\#8), and \textbf{corrections} of \textbf{factual errors} (\#8):

\#5 SPON: \textit{``I find it very good that parents are reminded about that. All parents should read this article! [...]''}

\#6 SPON: \textit{``Mr. Fleischhauer, what do the colleagues say about your comment? [...] Are you insane?''}

\#7 OMP: \textit{``One should not forget in an article like this to mention who's really to blame [...]''}

\#8 OMP: \textit{``[...] The author of this short note (either APA or Standard) has obviously very poor geography skills: the Traunstein is a very distinctive mountain in Austria [...]''}

\subsection{Comments Addressing the Moderator}
The authors of the following meta-comments complain and ask the moderator for the \textbf{rationale behind blocking} previous comments (\#9,\#10,\#12). One user \textbf{requests} a feedback \textbf{feature} for moderators so that users understand the rationale behind their decisions (\#9,\#11):

\#9 SPON: \textit{``[...] It would be beneficial, if you could receive brief feedback on the censored contributions, why the censorship occurred. If e.g. in a longer post a part does not conform to the guidelines, one could replace it with a ``[because of xxx]'', where instead of xxx it says ``insulting other participants'' or ``glorification of violence'' or whatever. A few template formulations would be enough. Then one would at least know why a contribution was censored and could be addressed in future contributions.''}

\#10 SPON: \textit{``It seems as if postings with the reference to ``censorship'' were systematically deleted here in the forum. Would you like us to spread this fact in other forums, blogs, etc.? Where among other things has this post remained: [link to a screenshot] Nothing against a deletion of unclean and unlawful contributions. [...]''}

\#11 OMP: \textit{``Uiui, Standard deletes already published comments. I would like to know how...''}

\#12 OMP: \textit{``Haha and DER STANDARD actually censored a posting from me again. Why? [...]''}

\section{Threats to Validity and Limitations}
\label{sec:threats}
We mention limitations to its internal and external validity. Regarding the internal validity, this study contains multiple coding tasks, and human coders can cause noise in the training set data. We dealt with that issue, by designing a coding guide over many iterations \cite{neuendorf2016content}. It defines the criteria for a comment to belong to a specific meta-addressee class with examples. However, annotating 1,000 random user comments is tedious. Some user comments are long, and the comment classes occur at imbalanced frequencies. For example, the internal media responsibilities are unclear, whereby the coders sometimes assumed the addressee. For example, SPON uses the username ``sysop'' to reply to single user questions, but it is unclear who composes these comments. This uncertainty caused disagreements between the peer-coders. 

Addressees in comments is a broad field and users also address and mention, for instance, celebrities, institutions, other users, or the general public. This study only focuses on the identification and classification of German meta-comments. However, it is possible to categorize meta-comments into a different set of addressee-classes which would lead to different results. We sampled part of our SPON training set based on regular expressions due to the small share of meta-comments. This procedure affected the ANOVA F-value as well as the significance of word-based features for the SPON dataset.

Regarding external validity, our work uses comments from the news sites SPIEGEL Online and DER STANDARD. User comments posted on respective Facebook or Twitter pages might use different terms or have a different style of writing. The accuracy of our classifier might be different.

The cross-dataset classification in Table \ref{tab:cross_data_source_results} is an initial step to check whether the automatic classification can be used for comments on other media companies' sites without using labeled data from their site. When training the traditional classifier on the OMP dataset and testing it on the SPON comments, we achieved a promising $F_{0.5}$ score of 0.85. However, as we used user comments from only two different datasets, further evaluation will be needed in the future if we are to generalize this statement.

\section{Related Work}
\label{sec:rel-work}
The question of who is addressed in user comments has been tackled in different studies, by different means, and for different purposes. We are currently carrying out a systematic literature review, covering the state of current research on the content analysis of user comments in online news media. To date, we have found related works that consider the variety of addressees of user comments. Most of these works conducted a qualitative content analysis and manually identified the addressees.

Collins and Nerlich \cite{collins2015examining} manually labeled direct references to other users and to the author to investigate public deliberation. Gervais \cite{gervais2015incivility} studied incivility in online user comments. Bergt and Welker \cite{bergt2013online} conducted a manual content analysis of 4,840 German user comments to check whether users refer to the quality criteria of news coverage and how it is integrated. They found that 5.9\% of user comments refer to quality criteria. Lopez-Gonzalez and Guerrero-Sole \cite{lopez2014medium} carried out a manual content analysis to analyze how much hate speech users direct towards the medium. They found that 2.84\% of comments address the medium.

Macovei \cite{macovei2013neo} conducted a case study and manually analyzed 1,000 Romanian reader comments on articles about a protest. In this respect, she qualitatively analyzed the users' expressions towards the newspaper, the authors, or to other users. Manosevitch and Walker \cite{manosevitch2009reader} analyzed the potential of the readers' comments section as a constructive space for public discourse. In this regard, they manually analyzed the social process of deliberation of 124 comments where they identified how users address other users, post questions, and address an article's content.

Rowe \cite{rowe2015deliberation} explores the differences in deliberative quality between news website users and Facebook users. To measure interactivity, he manually labeled comments that refer to other users. Al-Rawi \cite{al2017assessing} also analyzes the sentiment of Facebook comments. He studies the most recurrent words and phrases to assess the overall sentiment towards the topics being addressed. Carvalho et al.~\cite{carvalho2011liars} have analyzed comments on political debates, in which they manually identified ``opinion targets''. Opinion targets can be politicians, relevant media personalities, or other commentators. These can be politicians participating in the televised debates or other relevant media personalities. Further, they manually annotated how human entities are mentioned in user comments, for instance, by name, position, or nickname. Word embeddings capture this automatically.

Park et al.~\cite{park2016supporting} developed a system for supporting comment moderators that identifies high-quality comments by using different analytic scores. One feature is based on the LIWC dictionary to measure users' personal experiences. Instead of measuring quality from the users' perspective, we focused on identifying meta-comments, with a supervised learning approach. Djuric et al.~\cite{djuric2015hate} have utilized comment embeddings with paragraph2Vec to classify hate speech in comments.

Schabus et al. \cite{Schabus2017} created the OMP dataset, which contained annotated comments for different categories. In our work, we reused the ``Feedback'' category as meta-comments and were able to outperform their classification results. Fast et al. \cite{fast2016empath} and Park et al. \cite{park2018conceptvector} developed a prototype that analyzes user comments with respect to concepts. Their prototype uses word embeddings to extend the keywords given by the user to generalize a concept. Hullman et al. \cite{hullman2015content} conducted a qualitative content analysis of user comments on presented visualizations and found that over one third of the analyzed comments provided direct critical feedback on the journalistic content. They also suggest improving the design of commenting interfaces by grouping user comments according to their reference. Google and Jigsaw have established a project called Perspective \cite{perspective_2018} that uses machine learning to automatically detect toxic language in user comments. They published an experimental model that identifies attacks on the article's author in user comments which is a subset according to our meta-comment definition. To the best of our knowledge, we did not find any other work that presents an automatic approach for the identification and classification of meta-comments.

\section{Discussion}
\label{sec:discuss}


This paper focuses on automatically identifying and classifying meta-comments -- while maximizing the accuracy and generalizability of the automated approach. Our classification approach was inspired by previous work by Maalej and Nabil \cite{maalej2015bug} who classified app reviews in the domain of mobile app stores into four different feedback categories. We discuss the findings from both the technical and the application perspectives. 

\subsection*{Using and Improving the Approach on Different Datasets}
We expect our supervised learning approach to be applicable to other comment sections and other languages as it only requires the comment text and basic metadata. Applying our approach to other languages would require as many user comments as possible to precisely capture word similarities with word embeddings in that language. Additionally, a training set of a similar size to ours would be needed. The remainder of the process is language independent. One advantage of our approach is that it operates without common natural language processing methods such as lemmatization, named entity recognition, or part-of-speech tagging, which depend on pre-trained language specific models. Although word embeddings are also language specific, we can train them unsupervised on a large corpus of user comments to find words that users use in a similar context. However, it is unclear whether our approach is generalizable in other domains, for example, as part of online courses where students' comments might address teaching materials, instructors, forum-moderators, or other students; or an online store where users' comments might address vendors, developers, or delivery services.


We used transfer learning \cite{michalski1983theory} in the end-to-end classification by pre-initializing the embedding layer with pre-trained weights from the word embeddings. This approach did not use any hand-crafted features and achieved encouraging results with $F_{0.5}$ scores of 0.73 to 0.86. Typically, neural networks need large training sets to outperform traditional approaches \cite{Goodfellow-et-al-2016}. 
Traditional approaches often perform better on small training sets as domain experts implicitly incorporate significant information through hand-crafted features \cite{chollet2018deep}. 
We assume that for our experiments the hand-crafted keywords for the SPON dataset provided a considerable advantage whereas the end-to-end approach has to derive high-level features with many training samples. We presume that, given more training data, an end-to-end classification would outperform traditional approaches. 
More sophisticated features from the comment thread, comment ratings, user profiles, user comment history, or the respective article might improve the accuracy but this would require additional metadata from the comment section. 


\subsection*{Application and Utilization of User Feedback}
While this work is empirical and exploratory in nature, our intermediate goal is to develop and evaluate a tool for user comment analysis that we plan to evaluate with domain experts in future work. Our qualitative insights into identified meta-comments showed that our classification can capture meta-comments with diverse constructive feedback. A comment analysis tool can aggregate and forward the identified meta-comments to the concerned stakeholders. Further, it can enable moderators and journalists to directly reply to users to allow direct participation in the forum conversations while reducing the effort of manually searching for response worthy user comments.

\textbf{Media houses} can utilize user feedback from the  meta-comments. The commenters addressing the media houses demand a transparent prioritization of topics by the news. They further seek for understanding of journalistic production routines and the sources used for an online article. To meet this demand, media houses might aim to explain newsroom working routines.
An article recommendation system could utilize user recommendations as an input to highlight articles for other user groups. 
\textbf{Journalists} could reply to questions and aggregate frequent questions to a ``frequently asked questions'' section. Journalists could incorporate additional information provided by users either into the article or link to them. A new perspective might inspire journalists to produce an additional news article. Identifying meta-comments could help journalists to double check factual errors and fix them immediately.

In comments addressing the \textbf{moderator} users actively ask for the rationale behind blocking their comments. Users even show interest in improving their contribution if moderators would provide feedback about their decision. Forum moderators could reply to deescalate the dialog with unruly users. The online forum development team could consider user feature requests. For instance, a reply function for forum moderators to educate and provide feedback to users about what constitutes a desirable high-quality contribution. The dialogue between users and moderators could further help to improve the netiquette for user contributions.


Our classification approach is able to identify meta-comments that stakeholders deem useful, as they contain diverse user feedback and complaints, corrections, additional information, open questions, or clarification and feature requests. Feedback information of meta-comments could be further classified and clustered into categories, for example, as bug reports regarding the article, questions to the author, or forum feature requests. Subsequently, such automatic classification could help forwarding user comments to the relevant person responsible. In summary, identifying meta-comments would support stakeholders in extracting valuable information from user comments while also representing a crucial prerequisite for fostering a better dialog between media providers and users and increase the chances that response-worthy user comments are found at all.

\section{Conclusion}
\label{sec:conclusion}
With the emergence of user comments in online news media, news organizations are in need of tools to cope with the number of user comments. Researchers have found that journalists appreciate user feedback that, for instance, reports errors in articles, include additional information on a topic, or contain critique addressed to the quality of an article. In this paper, we present a preliminary approach to automatically identify and classify comments not (only) related to the news article but comments that address, for instance, the media company, a journalist, or a community-moderator. We call these comments ``meta-comments''.

By using a supervised machine learning approach, we achieved encouraging results with $F_{0.5}$ scores between 76\% and 91\%. We found similarities between the most significant features of 2 large datasets. We computed word and comment embeddings based on ~11 million German user comments for enriching text features, deriving semantic features, and transfer learning. The end-to-end learning approach outperformed the traditional approach on the ``One Million Posts'' dataset. We gained further qualitative insights into the content of automatically identified meta-comments. Finally, in our discussion, we highlight the training of word embedding models based on user comments as an important step for applying our approach to other languages. We further discuss use-cases for stakeholders, as e.g. considering the users' forum feature requests when further developing the news comment section.

\section*{Acknowledgements}
We thank V. Biryuk, J. Hennings, and H. Immler for their support with the manual labeling of the collected German user comments.

\bibliographystyle{ACM-Reference-Format}
\bibliography{bibliography}

\end{document}